\documentclass[preprint,aps,pra,superscriptaddress,groupedaddress]{revtex4}
\usepackage{epsfig}
\usepackage{hyperref}
\usepackage{braket}
\usepackage{bm,color}
\usepackage{amssymb}
\usepackage{enumitem}
\usepackage{float,CJK}
\usepackage{appendix}

\makeatletter
\@addtoreset{equation}{section}
\@addtoreset{figure}{section}
\@addtoreset{table}{section}
\makeatother

\def\be{\begin{equation}}
\def\ee{\end{equation}}
\def\ba{\begin{eqnarray}}
\def\ea{\end{eqnarray}}

\newcommand\hdot[1][2]{\mathbin{\vcenter{\hbox{\scalebox{#1}{$\bullet$}}}}}
\newcommand\hcir[1][2]{\mathbin{\vcenter{\hbox{\scalebox{#1}{$\circ$}}}}}
\begin{document}
\title{Board Games for Quantum Computers}
\begin{CJK}{UTF8}{gbsn}
\author{Biao Wu(吴飙)}
%\email{wubiao@pku.edu.cn}
\affiliation{International Center for Quantum Materials, School of Physics, 
Peking University, 100871, Beijing, China}
\affiliation{Wilczek Quantum Center, School of Physics and Astronomy, 
Shanghai Jiao Tong University, Shanghai 200240, China}
\affiliation{Collaborative Innovation Center of Quantum Matter, Beijing 100871,  China}
\author{Hanbo Chen(陈汉博)}
\affiliation{International Center for Quantum Materials, School of Physics, 
Peking University, 100871, Beijing, China}
\author{Zhikang Luo(罗智康)}
\affiliation{College of Chemistry and Molecular Engineering, Peking University, 100871, Beijing, China}
\date{\today}

\begin{abstract}
Scalable board games,  including Five in a Row (or gomoku) and weiqi (or go), are generalized 
so that they can be played on or by quantum computers. 
We adopt three principles for  the generalization: the first two are to ensure that the games 
are compatible with quantum computer and the third is to ensure that the standard classical games
are the special cases. We demonstrate how to construct basic quantum moves and  use them to set up 
quantum games. There are three different schemes to play the quantized games: one quantum computer
with another quantum computer (QwQ), two classical computer playing with each other on one 
quantum computer (CQC),  and one classical computer with another classical computer(CwC).  
We illustrate these results with the games of Five in a Row and weiqi. 
\end{abstract}

\maketitle
\section{Introduction}
\begin{figure}[h]
 \includegraphics[width=6cm]{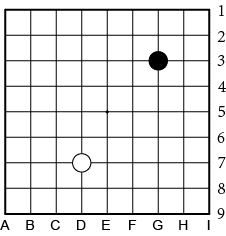}
        \caption{Scalable board. The board is  labelled horizontally by Roman letters and vertically 
        by Arabic numbers so that each point (or intersection) is specified by coordinates. 
        For example, the coordinates of the central point are $(E,5)$. 
        The game is played by two players who take turn to place black and white stones 
        on the points of the board. This board has a grid of 9$\times$9.  Even though the usual boards used in Five in a Row (FIR) and weiqi have
        grids of  15$\times$15 and 19$\times$19, respectively,  both games can be played on boards of other sizes.}
	\label{board}
\end{figure}

There are many interesting board games. Chess, weiqi (or go as usually called in English), and 
Five in a Row (or gomoku) are among the most popular. In particular, both weiqi and FIR (Five in a Row) are played 
on scalable square boards (see Fig. \ref{board}),  
where two players take turn to place black and white stones on the board. Following the convention  of 
quantum information,  we call the two players  Alice and Bob who play, respectively,
white stones and  black stones. In this work, we focus our attention on these scalable 
board games and discuss how to generalize them so that they can be played by or on quantum computers 
with intrinsic quantum moves.  

The most popular model of quantum computer consists of a set of quantum logical gates operating on 
an array of quantum bits (or qubits)~\cite{ChuangBook}.  
As there are three states, black stone, white stone, and unoccupied, at points (intersections of grid lines), 
the board, such as the one in Fig. \ref{board}, can be considered as an array of quantum trits  (or qutrits). 
It is therefore natural and convenient for us to use a different model of quantum computer, where a set of 
quantum logical gates operate on an array of qutrits~\cite{Qutrit}. 

We will begin  by considering the principles for designing scalable board games for quantum computers.  
First of all, the moves should be unitary and made of a number of quantum gates. 
Moreover,  as the board is scalable,  the number of quantum gates in a move should not be exponentially large. 
We  review the basic moves in the classical games of FIR and weiqi and show that they can be represented by unitary 
transformations, i.e., quantum gates. We then construct various quantum moves and use them to design games
for quantum computer.  Due to its simplicity, we use FIR as an example to discuss three different schemes to play 
quantum game: one quantum computer with another quantum computer (QwQ); two classical computers playing with each other
on a quantum computer (CQC), and one classical computer with another classical computer (CwC).   For the third scheme, 
as a classical computer can not store faithfully a quantum state in a Hilbert space of exponentially large dimension, 
one may have to impose an upper limit on the number of  superposed games.  The game of weiqi is much more complicated. 
We find that its quantum version may never finish when played by or on quantum computer.

% games are equivalent
% to physical systems with Ising spin-1  on two-dimensional square lattices. 
% When these games are played on quantum computers, these spins become quantum. 

\section{Principles of quantum games}
In the circuit model of quantum computer, there are a set of universal quantum gates that operate on an 
array of qubits or qutrits~\cite{ChuangBook,Qutrit}.  These quantum gates have two important features: (1) 
they are unitary transformations and (2) they involve only one or two qubits (or qutrits). 
In addition, when Hadamard gate and phase-related gates are not used,  a quantum computer becomes a reversible
classical computer.  To respect these three features, we should adhere to the following three principles for designing 
scalable board games played on or by quantum computers. 
\begin{itemize}
\item[{\bf P1.}] All the moves are unitary transformations;
\item[{\bf P2.}] All the moves involve only finite number of correlated quantum bits;
\item[{\bf P3.}] When it is limited to a true subset of all the moves, the game becomes classical. 
\end{itemize}

For a game to be played on or by quantum computers, 
moves must be quantum gates or combinations of various quantum gates. 
Since all the quantum gates are unitary transformations, all the moves must be unitary.
To explain the principle {\bf P2}, let us suppose there is  a move that involves $m$ 
correlated qubits and the number $m$ is proportional to  $n$, the total number of points on the board.   This move corresponds 
to a generic $3^m\times 3^m$ unitary matrix, which is usually  a product of  about $3^m$ to $3^{2m}$ basic 
quantum gates upon decomposition~\cite{ChuangBook}.  This implies that it would take an exponentially 
long time to execute a single move of this kind. This is not reasonable and so we have the principle {\bf P2}. 
The word ``correlated" is crucial here. For a move that involves  $m$  uncorrelated quantum bits, 
it can be expressed as a combination of $m$ one-qubit quantum gates (see, for example, the capture move in Section \ref{cmove}). 
The principle {\bf P2} is also related  to the $k$-local property of Hamiltonians that are used to model 
quantum computers~\cite{Dam}. The principle {\bf P3} is obvious. In this work, we only consider quantum 
board games that are generalized from the well-known classical games. In such cases, the principle {\bf P3} 
means that we recover the standard classical games when it is limited to a true subset of all the moves. 
It would be interesting if one can design a new quantum game that is not connected to any known classical game. 
In this case, when the principle {\bf P3} is applied to this new quantum game, we would have a new classical game.  

Just like in any classical game, the rules of a quantum  game should be fair to all players, for example, the player 
making the first move should have as little advantage as possible. Since this is not unique to quantum games, 
this is not listed in the above principles.

%%%%%%%%%%%%%%%%%%%%%%%%%%
\section{Review of classical games}
\label{cmove}
%%%%%%%%%%%%%%%%%%%%%%%%%%
We now review the moves in the classical games of FIR and weiqi. We show that all of them are special cases of 
quantum moves and obey the principles {\bf P1} and {\bf P2}.  
We use Dirac notations to denote the state at each point and the whole board configuration. 
For example, $\ket{\hdot}$, $\ket{\hcir}$, and $\ket{U}$ denote, respectively,  
a black stone, a white stone, and unoccupied at a given point.  For the whole board, 
one should specify the states for all the points on the board. However, for simplicity, we will not  
specify the unoccupied points explicitly in a ket (or bra) state.  For instance,  the board configuration in Fig.\ref{board} 
can be denoted as $\ket{\hdot_{G3}\hcir_{D7}}$, indicating  that there are a black stone at the point $(G, 3)$ and 
a white stone at the point $(D, 7)$ while no stones elsewhere. 
From now on, only in some special cases we will specify unoccupied points explicitly in ket (or bra) states. 

In the standard games of FIR and weiqi, there is one basic move, placing a black or white stone  on an unoccupied point.
Placing a black stone and a white stone on an unoccupied point can be mathematically expressed as
\be
\ket{\hdot}=X^b\ket{U}~,~~~~\ket{\hcir}=X^w\ket{U}\,,
\ee
where $X^b$ and  $X^w$ are  transformations that can be put  into matrix forms. If we let 
\be
\ket{\hdot}=\pmatrix{1\cr 0\cr 0}~,~~~~\ket{U}=\pmatrix{0\cr 1\cr 0}~,~~~~\ket{\hcir}=\pmatrix{0\cr 0\cr 1}~,~~~~
\ee
we have
\be
X^b=\pmatrix{0 & 1 & 0\cr 1& 0 &0\cr 0& 0 &1}~,~~~X^w=\pmatrix{1 & 0 & 0\cr 0& 0 &1\cr 0& 1 &0}
\ee
It is interesting and important to note that $X^{b}$ and  $X^{w}$ also  remove stones from the board, i.e., 
\be
\ket{U}=X^b\ket{\hdot}~,~~~~\ket{U}=X^w\ket{\hcir}\,.
\ee
Both $X^{b}$ and  $X^{w}$ are clearly unitary transformations and belong to a family of one-qutrit quantum gates~\cite{Qutrit}. 

\begin{figure}[h]
 \includegraphics[width=9cm]{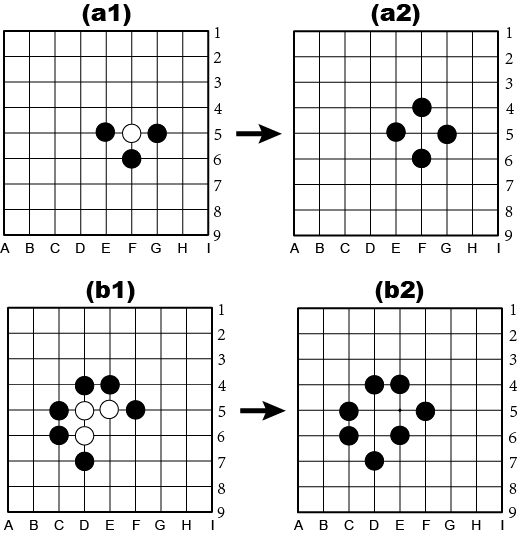}
        \caption{Two examples of capture in the game of weiqi. After one more black stone is placed on the board, 
        the white stones is(are) captured and removed from the board.}
	\label{capture}
\end{figure}

There is a special move in the game of weiqi, capture of stone(s).  Two examples are given in Fig.\ref{capture}. 
In the first example, after Bob places a black stone at the point $(E,6)$, the white stone at the point $(E,5)$ is captured 
and removed from the board. We can mathematically express the capture as  
\be
\ket{\hdot_{E5}\hdot_{F4}\hdot_{F6}\hdot_{G5}}=X^w_{F5}X^b_{F4}\ket{\hdot_{E5}\hdot_{F6}\hdot_{G5}\hcir_{F5}}\,.
\ee 
Although Bob placed only one stone on the board, due to the mandatory removal of the captured white stone, this  move 
consists of two one-qutrit quantum gates, $X^b_{F4}$ and $X^w_{F5}$. In many situations, multiple stones can be captured 
simultaneously. For the second example in Fig.\ref{capture},  after Bob places one black stone at the point $(E,6)$, 
three white stones are captured and removed. Mathematically, it can be expressed as 
\ba
&&\ket{\hdot_{D4}\hdot_{E4}\hdot_{C5}\hdot_{F5}\hdot_{C6}\hdot_{E6}\hdot_{D7}}\nonumber\\
&=&X^w_{D5}X^w_{E5}X^w_{D6}X^b_{E6}
\ket{\hdot_{D4}\hdot_{E4}\hdot_{C5}\hcir_{D5}\hcir_{E5}\hdot_{F5}\hdot_{C6}\hcir_{D6}\hdot_{D7}}\,.
\ea
In this case, one capture moves consist of four one-qutrit quantum gates. In the most extreme case where one player's stones 
are all captured at the late stage of the game, we would have a capture move consisting of a number of one-qutrit 
quantum gates that scales linearly with $n$, the board size.  

Overall, we see that all the moves in the classical games of FIR and weiqi obey the principles 
${\bf P1}$ and ${\bf P2}$. These classical moves will be a true subset of all the moves in the quantum games. 
%%%%%%%%%%%%%%%%%%%%%%%%%%
\section{Quantum moves}
\label{qmoves}
%%%%%%%%%%%%%%%%%%%%%%%%%%
In each of all the  classical moves,  the player places one stone at exactly one point. 
We now construct quantum moves where the player can place one stone simultaneously 
at different points.  As a result, one can view a quantum game as a superposition of finite number  classical games, 
which are played simultaneously. 
There are infinitely many quantum moves that satisfy the principles {\bf P1} and {\bf P2}. 
For simplicity, we will show only details of how to construct 
two types of superposition moves and three types of entangled moves. Construction of other quantum 
moves can be followed straightforwardly. 

Hadamard gate is the only one among the universal quantum gates that transforms a 
non-superposition state to a superposition state~\cite{ChuangBook}. A quantum computer with no Hadamard gate is 
essentially a  reversible  classical computer. So, to construct quantum moves, we have to use ternary extension 
of Hadamard gate. There are three of them~\cite{Qutrit}; we use only two. They are 
\be
H^b=\frac{1}{\sqrt{2}}\pmatrix{1 & 1 & 0\cr 1& -1 &0\cr 0& 0 &\sqrt{2}}~,~~~
H^w=\frac{1}{\sqrt{2}}\pmatrix{\sqrt{2} & 0 & 0\cr 0& -1 &1\cr 0& 1 &1}\,.
\ee  
These $H$ gates generate superpositions of a stone and unoccupied at a given point as follows
\ba
H^b\ket{\hdot}=\frac{1}{\sqrt{2}}(\ket{\hdot}+\ket{U})~,&&~~~~H^b\ket{U}=\frac{1}{\sqrt{2}}(\ket{\hdot}-\ket{U})\\
H^w\ket{\hcir}=\frac{1}{\sqrt{2}}(\ket{\hcir}+\ket{U})~,&&~~~~H^w\ket{U}=\frac{1}{\sqrt{2}}(\ket{\hcir}-\ket{U})\,.
\ea
%However, the ternary Hadamard gates $H^{b}$ and  $H^{w}$ themselves can not be used as  quantum move since
%they only produce half stone, not exactly one  stone, on the board. 
We will use them  with other gates to construct quantum moves. These ``other gates" include a pair of  
two-qutrit gates, $B^{(\alpha,i)}_{(\beta,j)}$ 
and $W^{(\alpha,i)}_{(\beta,j)}$, which are called controlled-X gates in quantum information~\cite{Qutrit}.  
If there is a stone at the point $(\alpha,i)$, $B^{(\alpha,i)}_{(\beta,j)}$ takes no action; if 
there is no stone at the point $(\alpha,i)$,  $B^{(\alpha,i)}_{(\beta,j)}$ takes an action  of $X^b$  at the point $(\beta,j)$. 
Similarly, 
if there is a stone at the point $(\alpha,i)$, $W^{(\alpha,i)}_{(\beta,j)}$ takes no action; if 
there is no stone at the point $(\alpha,i)$,  $W^{(\alpha,i)}_{(\beta,j)}$ takes an action  of $X^w$  at the point $(\beta,j)$.  
The matrix forms of  $B^{(\alpha,i)}_{(\beta,j)}$ and $W^{(\alpha,i)}_{(\beta,j)}$ can be found in Appendix A.

\begin{figure}[h]
 \includegraphics[width=9cm]{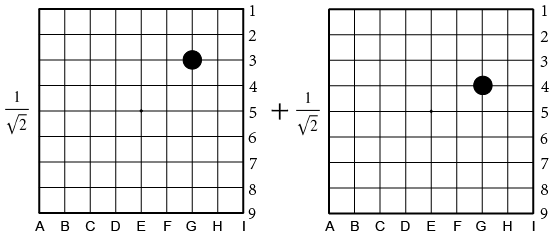}
        \caption{Superposition move. With $S^{b+}_{G3,G4}$, Bob places 
          one black stone  on the points $(G,3)$ and $(G,4)$ simultaneously with equal probability.}
	\label{super}
\end{figure}

With $H,B,$ and $W$ gates, we now introduce superposition moves.   A superposition move is to 
place one stone  simultaneously on two unoccupied points. 
Mathematically, for black stones, they are defined as
\be
S^{b+}_{(\alpha,i)(\beta,j)}=B^{(\alpha,i)}_{(\beta,j)}H^b_{(\alpha,i)}X^b_{(\alpha,i)}~~,~~~~
S^{b-}_{(\alpha,i)(\beta,j)}=B^{(\alpha,i)}_{(\beta,j)}H^b_{(\alpha,i)}\,.
\ee
One can check that 
\be
S^{b\pm}_{(\alpha,i)(\beta,j)}\ket{U_{\alpha,i}U_{\beta,j}}=
\frac{1}{\sqrt{2}}(\ket{\hdot_{\alpha,i}U_{\beta,j}}\pm\ket{U_{\alpha,i}\hdot_{\beta,j}})
\ee
We can similarly define $S^{w\pm}_{(\alpha,i)(\beta,j)}$ for white stones. One superposition move 
is illustrated in  Fig.\ref{super}.  If we let $\ket{\mathcal {U}}$ represent a board with no stones, 
the move in Fig.\ref{super} can be expressed mathematically as 
\be
S^{b+}_{G3,G4}\ket{\mathcal {U}}=\frac{1}{\sqrt{2}}(\ket{\hdot_{G3}}+\ket{\hdot_{G4}})\,.
\ee
After the superposition move by Bob, we  have two games playing at the same time.  This is an 
important feature  of superposition move: it doubles the number of games playing. 

When one player makes a move, the other player follows with a counter move. If the previous move is a one-qutrit move, i.e.,
$X^b$ or  $X^w$, there are two kinds of counter moves. The first kind is called one-to-one counter move 
$C^{(\alpha,i)}_{(\beta,j)}$, which  is essentially a two-qutrit gate: if there is a stone at the point $(\alpha,i)$,  then a stone of the opposite
color is placed at the point $(\beta,j)$; otherwise, no action is taken.   
The second kind is called one-to-two counter move $D^{(\alpha,i)\pm}_{(\beta_1,j_1)(\beta_2,j_2)}$:
if there is a stone at the point $(\alpha,i)$,  then a stone of the opposite
color is placed simultaneously at the points $(\beta_1,j_1)$ and $(\beta_2,j_2)$ with one of 
the four superposition moves, $S^{b\pm}_{(\beta_1,j_1)(\beta_2,j_2)}$ and $S^{w\pm}_{(\beta_1,j_1)(\beta_2,j_2)}$. 
It is clear that $D^{(\alpha,i)\pm}_{(\beta_1,j_1)(\beta_2,j_2)}$ are mathematically  three-qutrit gates. \\
 
\begin{figure}[h]
 \includegraphics[width=9cm]{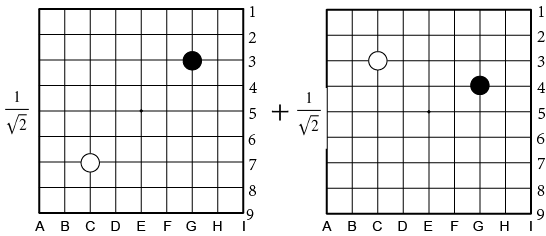}
        \caption{Entangled move:  one white stone is placed on two points simultaneously to counter  
        the  superposition move of the black stone in Fig.\ref{super}.}
	\label{entangle}
\end{figure}

An entangled move is a counter move for a   superposition move by the other player.  Using two counter moves, 
we can construct three different types of entangled moves. The first type involves two $C$ moves,  
\be
E^{(\alpha_1,i_1)-(\beta_1,j_1)}_{(\alpha_2,i_2)-(\beta_2,j_2)}=C^{(\alpha_1,i_1)}_{(\beta_1,j_1)}C^{(\alpha_2,i_2)}_{(\beta_2,j_2)}\,.
\ee
The second type involves one $C$ move and one $D$ move, 
\be
T^{(\alpha_1,i_1)-(\beta_1,j_1)}_{(\alpha_2,i_2)\pm(\beta_2,j_2)(\gamma_2,k_2)}
=C^{(\alpha_1,i_1)}_{(\beta_1,j_1)}D^{(\alpha_2,i_2)\pm}_{(\beta_2,j_2)(\gamma_2,k_2)}\,.
\ee
The third type involves two $D$ moves, 
\be
F^{(\alpha_1,i_1)\pm(\beta_1,j_1)(\gamma_1,k_1)}_{(\alpha_2,i_2)\pm(\beta_2,j_2)(\gamma_2,k_2)}
=D^{(\alpha_1,i_1)\pm}_{(\beta_1,j_1)(\gamma_1,k_1)}D^{(\alpha_2,i_2)\pm}_{(\beta_2,j_2)(\gamma_2,k_2)}\,.
\ee
The first type corresponds to a four-qutrit gate;  the second type a five-qutrit gate; the third type a six-qutrit gate. 
For example, for the superposition move in Fig.\ref{super}, we can counter an entangled move (see Fig.\ref{entangle})
\be
E^{G3,G4}_{C7,C3}\frac{1}{\sqrt{2}}(\ket{\hdot_{G3}}+\ket{\hdot_{G4}})
=\frac{1}{\sqrt{2}}(\ket{\hdot_{G3}\hcir_{C7}}+\ket{\hdot_{G4}\hcir_{C3}})
\ee
In general, there are more quantum moves, for example, using complex coefficients and involving more qutrits. 
These moves can all be constructed similarly and used in designing board games for quantum computers. We for simplicity 
have only discussed superposition moves and entangled moves introduced above.  It is clear that all these quantum moves
obey the principles {\bf P1} and {\bf P2}.

We end this section with an example of quantum moves that violate the principle {\bf P2}. 
Suppose after a move by Bob the wave function of the board becomes
\be
\ket{\Psi}=\sum_{j=1}^{M} a_j \ket{\phi_j}\,,
\label{bstate}
\ee
where $M$ is the total number of games going on and $\ket{\phi_j}$ is one of the games. If the best counter move 
for $\ket{\phi_j}$ is a white stone at the point $p_j$, then the best  that Alice can do is to 
place a white stone simultaneously at the points $p_j$'s, respectively,  for all the games $\ket{\phi_j}$'s. 
This is similar to entangled moves but at much larger scales. To do it,  Alice 
have to select a group of  points on the board, whose states are distinct for every $\ket{\phi_j}$,  and 
use them as the control qutrits to make the move.  If  this group has $G$ points, the total number of qutrits involved is around $G+N_M$, 
where  $N_M$ is the number of different $p_j$'s.  
In the late stage of the match, both $G$ and  $N_M$ 
should scale with the total number of points on the board.   As many of the qutrits are used as control qutrits, 
the involved qutrits are clearly correlated and thus violate the principle {\bf P2}. 
Since this kind of quantum move is made by considering distinct features 
of every game $\ket{\phi_j}$, we call it game-wise move. All the game-wise moves clearly violate the principle {\bf P2}. 
Since each $p_j$ is, respectively, the best move for $\ket{\phi_j}$, the principle {\bf P2} implies that 
a quantum computer is inherently incapable of executing the best move possible when the number of superposed games is 
exponentially large.

\section{Game  of FIR for quantum computer}
The basic rules of the classical game of FIR (Five in a Row) are as follows. Alice and Bob take
turn to place stones on a  15$\times$15 board with Bob, who has black stones, playing first. The stones can only be placed
on unoccupied points. The player who has managed first to have five stones of the same color  in a row (horizontal, vertical, or diagonal) 
wins. Only one-qutrit gates are used in this classical game. 

The basic rules of FIR for quantum computers are as follows. The board size is chosen to be the largest allowed by  
the quantum computer. Alice and Bob take turn to make moves with Bob playing first. 
There are now quantum  moves available, for example, superposition moves and entangled moves discussed above. 
As a result, there are more than one games going on in the perspective of classical players. For example, if Bob makes a first move
as in Fig. \ref{super} and Alice follows with an entangled move in Fig. \ref{entangle}. We have two games playing simultaneously. 
In general,  the state of the board  should have a form as in Eq.(\ref{bstate}) with  a total of $M$ games playing simultaneously. 
We define the rule of winning as follows. If there are five black stones in a row in any of the games $\ket{\phi_j}$, Bob wins; 
if five white stones in a row in any  $\ket{\phi_j}$, Alice wins. 

\begin{figure}[h]
 \includegraphics[width=7.5cm]{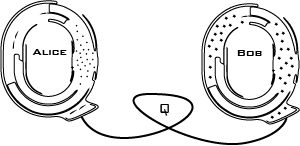}
        \caption{Quantum computer Alice plays a board game with another quantum computer Bob. 
        The state of the board game, which is a multiqutrit quantum  state, is transferred by a 
        quantum communication network between them. }
	\label{Q2Q}
\end{figure}

There are three different schemes to play quantum FIR. The first scheme, denoted as QwQ, is between two 
quantum computers Alice and Bob, who take turn to make moves.
The board state is stored on both quantum computers. After one player makes a move, the updated board state
is transferred by a quantum communication network to the other player. This  QwQ scheme 
is illustrated in Fig.\ref{Q2Q}.  There are at least two enormous challenges to realize this scheme for quantum games. 
The first challenge is of theoretical nature: it is not clear how a quantum computer itself can assess 
the quantum state of the board and decide what move to make next. The second is a technical challenge. 
So far, there are no universal fault-tolerant quantum computers and 
people have only realized the communication of the quantum state of a single qubit whereas the state of the board
is a multiqutrit quantum state.

\begin{figure}[h]
 \includegraphics[width=8cm]{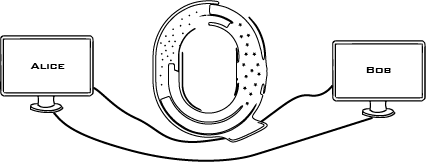}
        \caption{Classical computer Alice plays a quantum board game with another classical computer Bob on a quantum 
        computer. The two classical computers execute quantum gates in turn on the quantum computer and 
        update the state of the board. At the same time they communicate classically between them about the moves. }
	\label{CQC}
\end{figure}

The second scheme, called CQC,  is that two classical  computers Alice and Bob play a quantum game on a 
quantum computer (see Fig.\ref{CQC}).  The state of the board is stored on the quantum computer. 
Whenever, a classical computer, say, Alice, makes a move, she executes the corresponding quantum gates
on the quantum computer and updates the state of board. At the same time, she communicates her move, which is classical information, to 
Bob. In this scheme, there is also a theoretical challenge, how a classical computer assesses the quantum state of the board. 
The classical computer can accomplish this by measurement. However, this is destructive and, more importantly, 
it becomes impractical when the number of superposed games $M$ is exponentially large. In particular, one is interested  
to know whether there are five stones of the same color in a row in the board state and whether it is possible to do it without using destructive
measurement.  Note that this challenge is different from the one in  QwQ, where the 
quantum computer tries to assess its own state. 

\begin{figure}[h]
 \includegraphics[width=9cm]{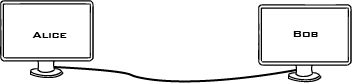}
        \caption{Classical computer Alice plays a quantum board game with another classical computer Bob. 
        In this case, the quantum state of board is stored classically on both  two classical computers. 
        Alice and Bob  communicate classically between them about the moves.  }
	\label{C2C}
\end{figure}

In principle, we can simulate any quantum system on a classical computer if the classical computer
has exponentially large amount of physical memory. Therefore, two classical computers, 
Alice and Bob, can also play a quantum board game. This scheme is called CwC and is shown in Fig.\ref{C2C}.  
However, since we do not have exponentially large memory in reality, we will have to limit the number of 
superposed games when the quantum game is played by two classical computers.  Suppose that the upper limit is $J$, 
which is independent of the board size. Then after some quantum moves by both players, 
the board state will reach the limit and become
\be
\ket{\Psi}=\sum_{j=1}^J b_j \ket{\phi_j}\,.
\label{limit}
\ee
All  these games $\ket{\phi_j}$ are stored on the classical computers along with $b_j$'s. After this limit, 
both Alice and Bob are only allowed to make classical moves or game-wise moves. Here the game-wise moves
are allowed because the number of games $J$ is fixed and does not scale with the board size. With this limitation, 
the scheme CwC is clearly feasible. 

Let us now re-consider the scheme CQC. It is possible that CQC is equivalent to CwC. 
Here is the explanation. 
In CQC, the board state $\ket{\Psi}$ on the quantum computer  is uniquely determined by the 
sequence of quantum moves made alternatively by Alice and Bob since  $\ket{\Psi}$ can be written as 
\be
\ket{\Psi}=Q^w_ZQ^b_{Z-1}\cdots Q^w_{j+1}Q^b_{j}\cdots Q^w_2Q^b_{1}\ket{\mathcal{U}}\,,
\ee
where $Q^w_{j+1}$ and $Q^b_j$ are quantum moves made, respectively, by Alice and Bob.  
This sequence of $Q^w$'s and $Q^b$'s are classical information but they uniquely determine 
the games $\ket{\phi_j}$ and their coefficients $a_j$ in Eq.(\ref{bstate}). 
Note that the number of $Q^w$'s and $Q^b$'s is only polynomially large while $M$, the number of games $\ket{\phi_j}$, 
can be exponentially large. This shows  an interesting fact  that a quantum state, which has too many components
to be stored directly on a classical compute, can still be coded as classical information of polynomial size. 
Both of Alice and Bob are aware of the sequence of $Q^w$ and $Q^b$. If they can 
extract information about the games $\ket{\phi_j}$ from $Q^w$ and $Q^b$ and decide what is the best next move, 
then we would no longer need the quantum computer to store $\ket{\phi_j}$ and their coefficients $a_j$. 
The two classical computers, Alice and Bob, in principle can play the quantum game by knowing just the sequence of $Q^w$ and $Q^b$.
As a result, CQC would become CwC. The problem at this moment is that it is not clear how one can  extract information 
about the games $\ket{\psi_j}$ from $Q^w$ and $Q^b$. One possibility is to use machine learning. 

The above discussion may appear general, not limited to the game of FIR. In a way, it is. However, 
as we will see in the next section, if the quantum game of weiqi is played
in both QwQ and CQC, it may never finish.  Such a possibility for FIR is very remote. 

\section{Game  of weiqi for quantum computer}
\begin{figure}[h]
 \includegraphics[width=7.5cm]{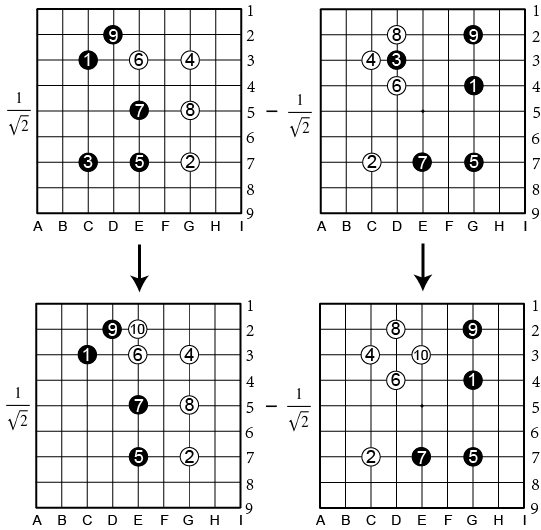}
        \caption{One possible way of capture a stone in the quantum game of weiqi: the captured stone is removed from all
        the games playing in parallel. The numbers on the stones mark the order that the stones are placed on the board: 
        the black stone with 1 is the first stone placed on the board, the white stone with 2 is the second placed, etc. }
	\label{qcapture1}
\end{figure}

\begin{figure}[h]
 \includegraphics[width=7.5cm]{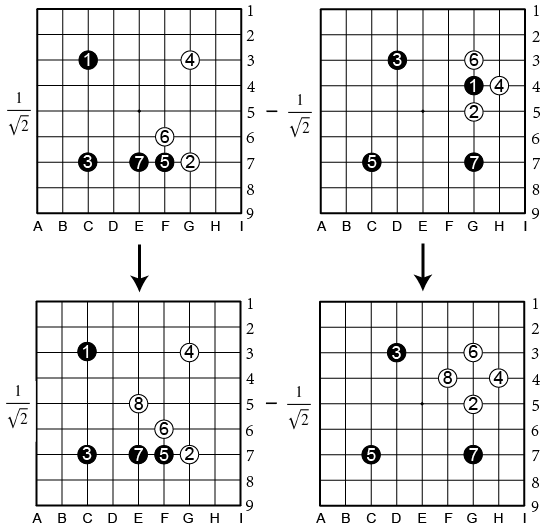}
        \caption{Another possible way of capture a stone in the quantum game of weiqi: the stone is removed from 
        the game where it is captured and stays in the other game. The numbers on the stones mark the order that the stones are placed on the board: 
        the black stone with 1 is the first stone placed on the board, the white stone with 2 is the second placed, etc. }
	\label{qcapture2}
\end{figure}

The key difference of weiqi from FIR is that there is capture of stones of the opposite color. 
To discuss  the quantum game of weiqi, we need to generalize capture quantum mechanically. 
However, this generalization  is highly non-trivial. There are at least two possible approaches. 
In the first approach, when a stone is captured in one game, it is removed
from all the games.  One example is shown Fig. \ref{qcapture1}, where the black stone with 3 is captured in the right game
and it is removed from both the left and right games.  In the second approach as illustrated in Fig. \ref{qcapture2},
the stone is removed only from the games where it is captured and stays in the games where it is still alive. 
For the first approach, if the stone was placed on the board by a classical move or a superposition move, 
it can be removed by  executing the inverse move. However, if the stone was placed with one of the 
entangled moves, its removal becomes difficult. As the stone to be removed has different surrounding 
in different games $\ket{\phi_j}$, thus the removal is a game-wise move. In the second approach, 
it is obvious that the removal of the stone is a game-wise move as one has to examine every game to
determine whether the stone should be removed.  Therefore, both approaches 
violate the  principle {\bf P2}. 

\begin{figure}[h]
 \includegraphics[width=7.5cm]{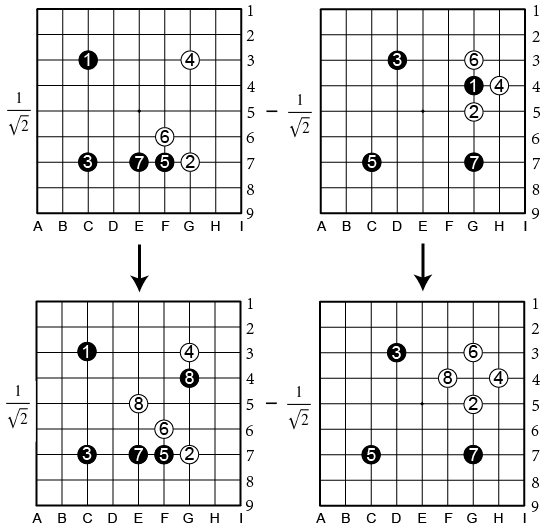}
        \caption{A legitimate way of capture a stone in the quantum game of weiqi: the stone at a certain position is removed from 
        the game when it is captured and a stone of the same color appears at the same position in the other game. 
        The numbers on the stones mark the order that the stones are placed on the board: 
        the black stone with 1 is the first stone placed on the board, the white stone with 2 is the second placed, etc. }
	\label{qcapture3}
\end{figure}

To satisfy the  principle {\bf P2}, we have to adopt an approach of capture  that appears odd but is fundamentally logical.  We use the example 
in Fig.\ref{qcapture3} to illustrate. Before the eighth stone, which is white, is placed on the board, the wave function of the board is
\be
\ket{\Phi_0}=\frac{1}{\sqrt{2}}\Big(\ket{\hdot_{C3}\hcir_{G7}\hdot_{C7}\hcir_{G3}\hdot_{F7}\hcir_{F6}\hdot_{E7}}-
\ket{\hdot_{G4}\hcir_{G5}\hdot_{D3}\hcir_{H4}\hdot_{C7}\hcir_{G3}\hdot_{G7}}\Big)
\ee
After the eighth stone is placed by Alice with an entangled move, the first stone, which is black, is captured in the game on the right side. 
This triggers a mandatory 
move, removing the first stone from the right game. Mathematically, this is given by $X^b_{G4}$. This means that, if we apply 
this mandatory move to every game, we have
$\ket{\Phi_0}$ as follows
\ba
\ket{\Phi_1}=X^b_{G4}E^{E7-E5}_{G7-F4}\ket{\Phi_0}
&=&\frac{1}{\sqrt{2}}\Big(\ket{\hdot_{C3}\hcir_{G7}\hdot_{C7}\hcir_{G3}\hdot_{F7}\hcir_{F6}\hdot_{E7}\hcir_{E5}\hdot_{G4}}\nonumber\\
&&-\ket{\hcir_{G5}\hdot_{D3}\hcir_{H4}\hdot_{C7}\hcir_{G3}\hdot_{G7}\hcir_{F4}}\Big)
\label{captureq}
\ea
The result is, of course, odd and surprising to people who are familiar to the game of weiqi: for the game on the left side, there is a black stone 
with even number and it is in a way placed there by Alice.  However, it is fundamentally logical and does not violate the principle {\bf P2}. 
There may be other ways to generalize capture to the quantum game of weiqi. This approach appears to the simplest and most straightforward. 
It is important to note the effects of $X^b_{G4}$: it removes it if there is a black stone at the point $(G,4)$;  
it places a black stone at the point $(G,4)$ if it is unoccupied; it has no effect if there is a white
stone at the point $(G,4)$. Its first two effects are illustrated in Fig.\ref{qcapture3}. The third effect means that 
if there is a white stone at the point $(G,4)$ in the left game, it stays there. 

\begin{figure}[h]
 \includegraphics[width=8cm]{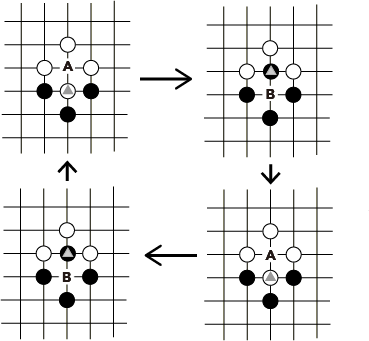}
        \caption{Ko rule for  unending cycle of captures. According to this rule,  
        after the capture of the marked white stone (upper left), Alice  is not allowed to place a white stone  on position B to capture 
        the marked black stone. She has to wait for at least one move. }
	\label{Jie}
\end{figure}

In the traditional game of weiqi, there could be unending cycles of captures by both players. One example is 
shown in Fig.\ref{Jie}: after  the marked white stone is captured by the black,  Alice
 can place a white stone at position B to capture the marked black stone; then the marked white stone can be captured again
 by Bob. To prevent this kind of fruitless unending cycle, there is the ko rule. For the situation in Fig.\ref{Jie}, 
 after the marked white stone is captured,  the ko rule forbids Alice to capture the marked black stone
 immediately, and Alice has to make at least one different move in other parts of the board before being allowed
 to capture the marked black stone. It is very unlikely that there are  unending cycles in the quantum game of weiqi for two reasons:
 (1) there are usually multiple games going in parallel; (2) the quantum capture is very different from its classical counterpart when 
 there are multiple games simultaneously. 

In the game of weiqi, there are forbidden points where a stone of a certain color is not allowed. For example, in the lower right game of 
Fig.\ref{qcapture3}, the point $(G,4)$ is forbidden for black stone.

We are now ready to state the basic rules of weiqi for quantum computers: Alice and Bob take turn to make either classical or quantum moves 
with Bob playing first. In each move, a stone can not be placed at forbidden points. When an unending cycle appears, the ko rule is enforced. 

Same as FIR, there are three schemes to play the quantum game of weiqi.  
However,  both the schemes QwQ and CQC are not very meaningful for weiqi. 
The reason is that the game of weiqi is much more complicated. In an ordinary classical game of weiqi, 
there are usually more than 200 moves near the end. This means that there are more than 200 stones on a 19$\times$19 board. 
Therefore, for the quantum game of weiqi, at the late stage of a match. we expect similarly that there are 
around 200 stones on the board and, in addition,  there are exponentially large number of games going on simultaneously. 
Suppose that Bob now wants to make a classical move, placing a stone at a given point. However, this given point at this stage 
is very unlikely  unoccupied for all the games $\ket{\psi_j}$. As a result, this move by Bob can remove a black stone in some 
games $\ket{\psi_j}$. It is well known in the game of weiqi, if the number of stones keep reducing, the game may never finish. 
To make legitimate moves in the late stage of the weiqi game, one has to use game-wise moves, which violate
the principle {\bf P2}.  There is also such a possibility for FIR. However, FIR usually does not last that long and 
will likely to end before there are too many stones on the board.  It would be very interesting  if one could  
design a new and competitive board game for quantum computer which has  no  shortcoming discussed above.  
 
For the game of weiqi, the scheme CwC is also feasible when we impose limitation on the number of superposed games. 
We have already noted in the discussion of FIR that game-wise moves are allowed in this scheme. 
It is similar here for the game of weiqi with an interesting twist: we  use either  
the quantum capture illustrated in Fig.\ref{qcapture1} or the one illustrated in Eq. (\ref{captureq}) and Fig.\ref{qcapture3}. 
We do not use the capture shown in Fig.\ref{qcapture2}. This is to keep the games in Eq. (\ref{limit}) related. 
In the game of FIR, when the limit is reached, it becomes clear that the best strategy is to examine 
every game $\ket{\phi_j}$ in Eq. (\ref{limit}), find its best move $p_j$,  and make a game-wise move, that is, placing
a stone simultaneously at points $p_j$'s, respectively, for games $\ket{\phi_j}$'s. As a result, the games $\ket{\phi_j}$
become independent of each other. In the game of weiqi, if we adopt the capture shown in Fig.\ref{qcapture2}, 
the games $\ket{\phi_j}$ would also become independent of each other. It is clearly more interesting to 
keep the games related. For this reason,  we  use either of the other two approaches of capture in the scheme CwC. 
In fact, we prefer the approach of Fig.\ref{qcapture1} since the one in Fig.\ref{qcapture3} may appear too odd
for traditional players.

%Suppose that the final state of the board  at the end of the games is
%\be
%\ket{\Psi}=\sum_{j=1}^M a_j \ket{\psi_j}
%\ee
%Let $\hat{C}$ be an operator such that $\hat{C}\ket{\psi_j}=C_j\ket{\psi_j}$ where $C_j$ is the number of points occupied by 
%black stones. We define
%\be
%\braket{\hat{C}}=\braket{\Psi|\hat{C}|\Psi}\,.
%\ee
%If the board size is $19\times 19$, Bob wins if $\braket{\hat{C}}\ge 185$ and Alice wins if $\braket{\hat{C}}< 185$.  In the classical game of weiqi, 
%Bob has advantages by playing first. To compensate that for Alice, Bob has to occupy  185 points or more to win. In the quantum game, 
%Bob may not have this much advantage or have even bigger advantage. In these cases, the number 185 should be adjusted accordingly. 

%%%%%%%%%%%%%%%%%%%%%%%%%
\section{Discussion and Conclusion}
%%%%%%%%%%%%%%%%%%%%%%%%%
We have discussed  the three principles for designing board games for quantum computer. The principle {\bf P2}
is particularly interesting and important as it exposes an inherent inability of a quantum computer to execute
the best possible move.  With two examples, 
Five in a Row (FIR) and weiqi, we have shown the details to construct quantum moves that obey these three principles. 
There have been some work discussing board games with quantum features~\cite{Qgo,Qchess}. 
The closest to our work is the discussion of  quantum chess in Ref.\cite{Qgo,Qchess}, where it is stressed 
that moves should be unitary. However, nothing similar to our principle {\bf P2} is discussed as chess 
is not a scalable board game. The generalized game of weiqi (or go)  in Ref.\cite{Qgo} has some
quantum features. But its moves are not always unitary and the generalized game can not be played on or 
by quantum computer. 

We have also discussed three different schemes to play the quantum games. The third scheme CwC is technically 
feasible and can be readily programmed. The other two schemes not only face daunting technical challenges 
as they involve yet-to-be-built universal quantum computer, but also need to overcome some theoretical issues. 
The future study of board games for quantum computer, in particular with the scheme QwQ, may lead to the understanding 
of how intelligence emerges out of the  quantum world. 

\acknowledgements
We dedicate this work to Professor P. W. Anderson, who was a very good amateur weiqi (or go) player and would 
probably view a quantum game of weiqi as a spin-1 system on a square lattice.   
This work is supported by the The National Key R\&D Program of China (Grants No.~2017YFA0303302, No.~2018YFA0305602), 
National Natural Science Foundation of China (Grant No. 11921005), and 
Shanghai Municipal Science and Technology Major Project (Grant No.2019SHZDZX01).

\appendix
\section{Matrix forms of controlled quantum gates $B^{(\alpha,i)}_{(\beta,j)}$ and $W^{(\alpha,i)}_{(\beta,j)}$}
Here we give the explicit matrix forms for quantum gates $B^{(\alpha,i)}_{(\beta,j)}$ and $W^{(\alpha,i)}_{(\beta,j)}$
introduced in Section \ref{qmoves}. If there is a stone at the point $(\alpha,i)$, $B^{(\alpha,i)}_{(\beta,j)}$ takes no action; if 
there is no stone at the point $(\alpha,i)$,  $B^{(\alpha,i)}_{(\beta,j)}$ takes an action  of $X^b$  at the point $(\beta,j)$. 
Following the conventional tensor  rule, we can write $B^{(\alpha,i)}_{(\beta,j)}$ as a $9\times 9$ matrix
\be
B^{(\alpha,i)}_{(\beta,j)}=\pmatrix{1&0&0&0&0&0&0&0&0\cr 0&1&0&0&0&0&0&0&0\cr 0&0&1&0&0&0&0&0&0
\cr 0&0&0&0&1&0&0&0&0\cr 0&0&0&1&0&0&0&0&0\cr 0&0&0&0&0&1&0&0&0\cr 0&0&0&0&0&0&1&0&0
\cr 0&0&0&0&0&0&0&1&0\cr 0&0&0&0&0&0&0&0&1}\,,
\ee
The move $W^{(\alpha,i)}_{(\beta,j)}$ is the counterpart for white stones and it has the following matrix form,
\be
W^{(\alpha,i)}_{(\beta,j)}=\pmatrix{1&0&0&0&0&0&0&0&0\cr 0&1&0&0&0&0&0&0&0\cr 0&0&1&0&0&0&0&0&0
\cr 0&0&0&1&0&0&0&0&0\cr 0&0&0&0&0&1&0&0&0\cr 0&0&0&0&1&0&0&0&0\cr 0&0&0&0&0&0&1&0&0
\cr 0&0&0&0&0&0&0&1&0\cr 0&0&0&0&0&0&0&0&1}\,.
\ee

\section{Absence of interference  in quantum board games}
Interference is a hallmark of quantum phenomena. In general, it can also exist 
in quantum board games.  However, in our quantum board games, if we limit ourselves to 
the classical moves, superposition and entangled moves introduced in Section \ref{qmoves}, interference  
does not exist.  Here is the proof.

Without loss of generality, we consider the state of the board after one of  Alice's moves
\be
\ket{\Psi}=a_1\ket{\phi_1}+a_2\ket{\phi_2}+\sum_{j=3}^{M}a_j\ket{\phi_j}\,.
\ee
Assume that interference happens here for the first time, say, $\ket{\phi_1}=\ket{\phi_2}$.  We will 
show that if this is the case, it would lead to  contradiction. Let 
\be
\ket{\tilde{\Psi}}=\sum_{j=3}^{\tilde{M}}\tilde{a}_j\ket{\tilde{\phi}_j}
\ee
be the state of board before the Alice's move. There must be at least two games $\ket{\tilde{\phi}_{j_1}}$ and $\ket{\tilde{\phi}_{j_2}}$, 
which become $\ket{\phi_1}$ and $\ket{\phi_2}$, respectively, after the Alice's move. 
That Alice's move only adds a white stone on the board has two implications: 
(1) the black stones in both $\ket{\tilde{\phi}_{j_1}}$ and $\ket{\tilde{\phi}_{j_2}}$
must be exactly the same; (2) the white stones are different. The former means that Alice's move can not be  an entangled move;
the latter implies that her move can not a classical move, either.

Alice's last choice is a superposition move that places a white stone to both positions $p_1$ and $p_2$. 
Assume that $\ket{\tilde{\phi}_{j_1}}$ turns to $\ket{\phi_1}$ with a white stone at $p_1$ and 
$\ket{\tilde{\phi}_{j_2}}$ turns to $\ket{\phi_2}$ with a white stone at $p_2$. 
With $\ket{\phi_1}=\ket{\phi_2}$, there must be a white stone at $p_2$ in the game $\ket{\tilde{\phi}_{j_1}}$ and
 a white stone at $p_1$ in the game $\ket{\tilde{\phi}_{j_2}}$. This means that
 the superposition move that  places a white stone to both positions $p_1$ and $p_2$ is not legitimate. Therefore, 
 Alice's last choice is  also impossible.

Based on the above discussion, we find that there is interference in our quantum board games 
if we limit ourselves to the classical moves, superposition moves and entangled moves.

\end{CJK}
%\bibliography{qboard}
\end{document}